\documentclass{aastex}
\usepackage{emulateapj5} 
\usepackage{graphicx}
\usepackage{times,psfig}

\def\ltsima{$\; \buildrel < \over \sim \;$}
\def\lsim{\lower.5ex\hbox{\ltsima}}
\def\gtsima{$\; \buildrel > \over \sim \;$}
\def\gsim{\lower.5ex\hbox{\gtsima}}
\def\mdot {\dot M}
\newcommand{\be}{\begin{equation}}
\newcommand{\en}{\end{equation}}
\newcommand{\ergs}{\rm \ erg \; s^{-1}}

\def\cmdue {\rm \ cm^{-2}}

\def\deg {^\circ}

\begin{document}

%apj
\received{~~} \accepted{~~}
\journalid{}{}
\articleid{}{}

\title{Swift observations of GRB\,050128: the early X--ray afterglow} 

\author{S.~Campana\altaffilmark{1}, L.A. Antonelli\altaffilmark{2},
G. Chincarini\altaffilmark{1,3}, S. Covino\altaffilmark{1},
G. Cusumano\altaffilmark{4}, D. Malesani\altaffilmark{5},
V. Mangano\altaffilmark{4}, A. Moretti\altaffilmark{1},
C. Pagani\altaffilmark{1,6}, P. Romano\altaffilmark{1},
G. Tagliaferri\altaffilmark{1},  M. Capalbi\altaffilmark{7},
M. Perri\altaffilmark{7}, P. Giommi\altaffilmark{7},
L. Angelini\altaffilmark{8,9}, P. Boyd\altaffilmark{8}, 
D.N. Burrows\altaffilmark{6}, J.E. Hill\altaffilmark{6}, 
C. Gronwall\altaffilmark{6}, J.A. Kennea\altaffilmark{6},
S. Kobayashi\altaffilmark{6,10}, P. Kumar\altaffilmark{11}, 
P. M\'esz\'aros\altaffilmark{6}, J.A. Nousek\altaffilmark{6}, 
P.W.A. Roming\altaffilmark{6}, B. Zhang\altaffilmark{12},
A.F. Abbey\altaffilmark{13}, A.P. Beardmore\altaffilmark{13},
A. Breeveld\altaffilmark{14}, M.R. Goad\altaffilmark{13}, 
O. Godet\altaffilmark{13}, K.O. Mason\altaffilmark{14}, 
J.P. Osborne\altaffilmark{13}, K.L. Page\altaffilmark{13}, 
T. Poole\altaffilmark{13} \& N. Gehrels\altaffilmark{8}  
}

\altaffiltext{1}{INAF -- Osservatorio Astronomico di Brera, Via Bianchi 46, I-23807
Merate (LC), Italy}

\altaffiltext{2}{INAF -- Osservatorio Astronomico di Roma, Via di Frascati 33,
I-00040 Monteporzio, Italy} 

\altaffiltext{3}{Universit\`a degli studi di Milano-Bicocca,
Dipartimento di Fisica, Piazza delle Scienze 3, I-20126 Milano, Italy}

\altaffiltext{4}{INAF -- Istituto di Astrofisica Spaziale e Fisica Cosmica Sezione di Palermo,
Via Ugo La Malfa 153, I-90146 Palermo, Italy}

\altaffiltext{5}{International School for Advanced Studies (SISSA-ISAS), Via
Beirut 2-4, I-34014 Trieste, Italy} 

\altaffiltext{6}{Department of Astronomy \& Astrophysics, 525 Davey Lab., Pennsylvania State
University, University Park, PA 16802, USA}

\altaffiltext{7}{ASI Science Data Center, Via Galileo Galilei, I-00044
Frascati (Roma), Italy}

\altaffiltext{8}{NASA/Goddard Space Flight Center, Greenbelt, MD 20771}

\altaffiltext{9}{Department of Physics and Astronomy, Johns Hopkins
University, 3400 North Charles Street, Baltimore, MD 21218}

\altaffiltext{10}{Center for Gravitational Wave Physics, Pennsylvania State
University, University Park, PA 16802, USA} 

\altaffiltext{11}{Department of Astronomy, University of Texas, RLM 15.308,
Austin, TX 78712-1083} 

\altaffiltext{12}{Department of Physics, University of Nevada, Box 454002, Las
Vegas, NV 891, USA}

\altaffiltext{13}{Department of Physics and Astronomy, University of Leicester, Leicester LE
1 7RH, UK}

\altaffiltext{14}{Mullard Space Science Laboratory, University College London,
Holmbury St. Mary, Dorking, RH5 6NT Surrey, UK}

\date{Received; accepted}
\email{campana@merate.mi.astro.it}

\begin{abstract}
Swift discovered GRB050128 with the Burst Alert Telescope and promptly
pointed its narrow field instruments to monitor the afterglow. X--ray
observations started 108 s after the trigger time. The early decay of the
afterglow is relatively flat with a temporal decay modeled with a power law
with index $\sim -0.3$. A steepening occurs at later times ($\sim 1500$ s) with
a power law index of $\sim -1.3$. During this transition, the observed X--ray
spectrum does not change. We interpret this behaviour as either an
early jet break or evidence for a transition from the fast cooling regime 
to the slow cooling regime in a wind environment. 

\keywords{gamma rays: bursts; X-rays: individual (GRB050128)}

\end{abstract}

\section{Introduction}

The Swift Gamma-ray Burst Explorer (Gehrels et al. 2004) was successfully
launched on 2004 November 20. Thanks to its fast-pointing capabilities, Swift
is performing the first comprehensive observations of the early afterglow
phase of Gamma--Ray Bursts (GRB). A few GRBs have been followed by Swift
within 200 s from their trigger time: GRB050117a (193 s), GRB050126 (131 s),
GRB050128 (108 s), GRB050215b (108 s), GRB050219a (92 s) and GRB050315 (83
s). 

In this paper we focus on GRB050128. The Burst Alert Telescope (BAT; Barthelmy
et al. 2005) on board Swift triggered and located GRB050128 at 04:19:54 UT
(Cummings et al. 2005). The burst profile is multi-peaked with a $T_{90}$
duration of 13.8 s. The fluence is $4.5\times 10^{-6}$ erg $\cmdue$ (15--350
keV) making it a `normal' burst with respect to the BATSE GRB population.
The spectrum of the burst during the $T_{90}$ interval can be described by a
power law model with photon index $\Gamma=1.5\pm0.1$ (15--350 keV). The peak
energy is above 350 keV making it a classical GRB.

Swift pointed autonomously to the GRB. We will report in the next sections
about the XRT and UVOT observations. 
Ground-based follow-up observations started as soon as the GCN circular
announcing the discovery of the new GRB was issued. This happened with some
delay, since Swift was in the early phases of the mission and each circular
was being checked manually before being distributed. The first GCN circular on
GRB050128 was issued by the XRT team (Antonelli et al. 2005). The robotic
60-cm REM telescope located in La Silla pointed to GRB050128 approximately 3
hr after the burst with good seeing conditions ($\sim 1''$). No new sources
were discovered with an upper limit of $H>17$ (Covino et al. 2005), $V>18.2$,
$R>18.2$ and $I>17.9$ (Melandri et al. 2005). A further upper limit came from
the 2-m Faulkes telescope South with $R>20.5$ 11.5 hr after the burst
(Monfardini et al. 2005). GRB050128 has also been observed in the radio band
at 8.4 GHz yielding an upper limit of 100 $\mu$Jy $\sim 11$ d after the burst
(Frail \& Soderberg 2005). 

In the following we focus on the observations by the X--Ray Telescope (XRT,
Burrows et al. 2005a) on board Swift. In section 2 we describe the data analysis. In
Section 3 we discuss on theoretical implications of these observations and 
in Section 4 we draw our conclusions. 

\section{XRT and UVOT observations}

UVOT (Roming et al. 2005) observations started on Jan 28, 2005 at 20:03:07 UT, 
about 15.8 hours after the burst. This delay happened because UVOT was in safe
mode at the trigger time. During each orbit UVOT acquired a single shot image
in each filter $U$, $B$ and $V$ in imaging mode. UVOT did not detect the
afterglow. Upper limits on the first (summed over about 1 d elapsed time) UVOT
observations are $V<18.9$ ($V<19.8$), $B<20.3$ ($B<21.1$) and $U<19.2$
($U<20.3$).    

XRT observed GRB050128 after an automatic slew of the Swift satellite. However,
XRT was not yet operating in its nominal automatic mode changing configuration
but rather in a manual mode for the purpose of obtaining calibration data.
Before pointing to GRB050128, the 
XRT was observing a UVOT calibration target in photon counting mode (see Hill
et al. 2004 for a description of XRT observing modes). XRT fully
settled on the BAT position 108 s after the trigger. XRT observed GRB050128
for 17 orbits following the first pointing, accumulating a nominal exposure
time of 17303 s (distributed over 73 ks). This low Earth orbit of Swift causes
source observations to be interrupted each orbit. At the same time, thanks
to the fast-pointing capability of Swift, several targets may be observed per
orbit.  
At this early stage of the mission the analysis of the data is not
straightforward. We analysed the data running the task {\tt xrtpipeline} within
FTOOLS v5.3.1 and cutting out temporal intervals when the CCD temperature was
higher than $-50\deg$ Celsius (see Burrows et al. 2005b) and when the total
count rate in the 0.2--10 keV energy band over the entire CCD was larger than
85 c s$^{-1}$ (these counts are mainly soft counts and are due either to dark
current or to the bright Earth limb near the end of each snapshot
observation). With these cuts we obtain a total exposure time of 13047 s
distributed over 11 orbits.   

\subsection{Angular analysis}

A fading source is clearly evident in all the XRT orbits. In the first two
orbits the source is clearly piled-up and to derive an unbiased position we
rely on the remaining $\sim 10$ ks exposure. An image has been extracted in
the 0.5--10 keV energy band to avoid contamination from low energy
photons. The source position has been derived with XIMAGE (v4.2.1) using the {\tt
centroid} command: RA(J2000): 14h38m18s.0, Dec(J2000):
-34$\deg$$45'$$55''$.9. The main contributors to the positional uncertainty
are the uncalibrated satellite attitude and boresight, resulting in a $\sim
6''$ error radius ($90\%$ confidence level).

\subsection{Temporal analysis} 

In order to properly track the decay of the fading source we have to account for
the piled-up core in the first two snapshot observations. To this aim we
extracted photons from an annular region (inner and outer radii 4 and 30
pixels, respectively) on source. This aperture was then applied to the rest of
the observations, even when it was not needed. The light curve will have an
underestimated normalization but it will not be distorted by pile-up.
A background light curve has also been extracted from an annular region (inner
and outer radii 80 and 120 pixels, respectively) centered on the same position
and free of other sources and hot pixels.
Fig. 1 shows the background-subtracted light curve in the 0.2--10 keV energy
band. The source is clearly fading. The decay light curve is not consistent
with a single power law ($\chi^2_{red}=2.6$, with 35 degrees of freedom, dof,
null hypothesis probability, nhp, $7\times 10^{-7}$),
being flat at the beginning and steepening at later times. We also tried a
single power law with a different initial time $t_0$, considered here as a
free parameter. We can account in principle for the observed decay with
$t_0=-780\pm290$ s, however this early time is not physically meaningful. A
better description of the data is provided by two power laws (with indices
$\alpha_1$ and $\alpha_2$) smoothly joined at a break time $t_b$ to model the
decay. The fit with this model is good ($\chi^2_{red}=0.7$ with 33 dof, $0.90$
nhp) and $\alpha_1=-0.27^{+0.10}_{-0.12}$, $\alpha_2=-1.30^{+0.18}_{-0.13}$
and $t_b=1472^{+300}_{-290}$ s (these errors are $90\%$ for one interesting
parameter, i.e. $\Delta\chi^2=2.71$, throughout the letter).   

\subsection{Spectral analysis} 

Given the large variability in the source count rate and the knowledge from the
temporal analysis of the existence of a change in the decay slope, we extracted
three spectra from our data, one from each of the first two snapshots and one
for the rest of the observation (see Fig. 1). The first two spectra were extracted 
from the same annular region as for the temporal analysis. The last spectrum, 
since the source is much fainter, was extracted from a circular region on source of 30
pixel radius. Exposure times are 286, 1653 and 10731 s, respectively.
Data have been filtered for grades 0--4 (according to the XMM-Newton
nomenclature, i.e. single and double pixel events\footnote{See e.g. the
XMM-Newton User's Handbook at
http://xmm.vilspa.esa.es/external/xmm\_user\_support/documentation/uhb/index.html}),
since at this stage the 
response matrix for the standard 0--12 pixel event is not fully calibrated. 
Ancillary response files were generated with the task {\tt xrtmkarf} within
FTOOLS (v5.3.1), accounting for the different extraction regions.
Data were rebinned to have at least 20 counts per energy bin and allow
$\chi^2$ fitting within XSPEC (v11.3.1). 

We fit the data with an absorbed power law model for all the observations. We
first fixed the absorbing column density to the Galactic value of
$N_H=4.8\times 10^{20}\cmdue$. We added a $5\%$ systematic uncertainty to all
our fits to account for the residuals still present in our response matrix
(given the relatively low number of counts this does not alter
our results sensibly). This simple fit can account for the observed
spectra. The fit is good with $\chi^2_{red}=1.1$ for 63 dof ($0.21$ nhp). The
power law photon index is $\Gamma=1.66^{+0.06}_{-0.07}$. The
0.2--10 keV unabsorbed fluxes of the three observations with mid times of 258
s, 6156 s and 51587 s are $2.2\times 10^{-10}$, $1.9\times 10^{-11}$ and
$6.6\times 10^{-13}\ergs\cmdue$, respectively. 
Given the slope change in the light curve we also untie the power law photon
index of the three observations (even if it is not required by the statistics).
The three photon indices are $1.59\pm0.08$, $1.79\pm0.11$ and $1.59\pm0.23$,
respectively. These values are consistent with each other with a small
deviation in the second snapshot. 
Although the fit is consistent with the Galactic column density, we let the
column density value free to vary. The fit is improved with a
$\chi^2_{red}=0.9$ (62 dof, $0.66$ nhp, see Fig. 2) and the improvement is significant
according to the F-test (probability $4\times 10^{-4}$, even if we improved an
already statistically good fit).
In Fig. 3, we show the contour plot of the column density vs. power law photon
index. The fit obtained with the column density fixed to the Galactic value is
outside the $3\,\sigma$ boundary. The absorbing column density is $(1.0\pm0.2)
\times 10^{21}\cmdue$ and the power law photon index $\Gamma=1.88\pm0.12$. 
Unabsorbed fluxes (0.2--10 keV) are $2.4\times 10^{-10}$, $2.0\times 10^{-11}$ and
$7.0\times 10^{-13}\ergs\cmdue$, respectively. Also in this case, leaving free
the photon index to vary within the observations the second one is
characterized by a slightly steeper index. 

\section{Physical interpretations}

The major result on the GRB050128 afterglow concerns the monitoring in the
X--ray band of its early temporal decay. This decay cannot be described by a
simple power law but can be accounted for by a slowly varying double power law
decay. During this transition there are no apparent marked spectral changes. 
The most straightforward interpretation is that the temporal break
reveals a jet, i.e. corresponding to the epoch when the relativistic
beaming angle ($1/\gamma$) becomes larger than the physical opening
angle ($\theta_j$) of the jet during the fireball deceleration
(e.g. Rhoads 1999).
In the slow cooling regime, for a uniform density circumstellar medium, the
temporal decay changes from $t^{3(1-p)/4}$ to $\sim t^{-p}$ (e.g. Rhoads
1999), which is well consistent with the observed temporal decay indices when
$p \sim 1.3$ is adopted. In such a case no spectral change is
expected. However, the expected spectral photon index should be $-(p+1)/2
\sim -1.15$, too small to be compared with the observed value. In order to
make the jet model work, one needs to assume $\nu_c < \nu_X < \nu_m$ before
the jet break, and $\nu_X > {\rm max}(\nu_c,\nu_m)$ after the jet break (here 
$\nu_m$ and $\nu_c$ are the typical synchrotron frequency and the
cooling frequency, respectively). 
In such a case, $p \sim 1.3$ gives a consistent interpretation of both
spectral and temporal indices in all three epochs, regardless of whether
the medium is an interstellar medium (ISM) or a wind from a massive
companion. 
This model requires a little bit of coincidence in that the synchrotron
frequency happens to cross the X--ray band during the jet break. However,
considering the rapid decline with time of $\nu_m$ this is not a very unlikely
possibility.
Another caveat is that a flat
electron spectrum $p \sim 1.3$ is abnormal in late afterglow fits
(e.g. Panaitescu \& Kumar 2001). However, since we are observing a  
previously unexplored early epoch, a small $p$ required for the jet model to
work cannot be ruled out. Possible ways to generate a flat electron
spectrum have been suggested earlier (e.g. Bykov \& M\'esz\'aros 1996). 
If this is indeed a jet break, this would be the earliest jet break
detection so far. 
Using the standard definition of jet break time (i.e. $\theta_j=1/\gamma(t_j)$)
one can derive $\theta_0=1.8\deg
(t_j/2000~{\rm s})^{3/8}\, (E_{52}/n)^{-1/8}\,[(1+z)/2]^{-3/8}$  for a
constant density interstellar medium and
$\theta_0=3.7\deg (t_j/2000~{\rm s})^{1/4}\, (E_{52}/A_*)^{-1/4}\,
[(1+z)/2]^{-1/4}$ for a wind model. Here $n$ is the density of the ISM,
$E_{52}$ is the isotropic-equivalent burst energy in units of $10^{52}$ erg
and $A=\mdot/(4\,\pi\,v)$ is the wind parameter, with $\mdot$ being
the mass loss rate, $v$ being the wind velocity, and $A_*=A/(5\times
10^{11}$ g cm$^{-1}$). 
These jets are not extremely narrow (e.g. Covino et al. 2003) but are narrower
than the typical jets identified in the previous late afterglow observations 
(e.g. Table 2 of Bloom et al. 2003). According
to the GRB standard energy argument (e.g. Frail et al. 2001; 
Panaitescu \& Kumar 2001), such a narrow jet should correspond to
large isotropic gamma-ray energy. Since this burst was not particularly
bright, it might lie in the low energy tail of GRB-energy distribution, thus
being another outlier for the standard energy relation. 

Besides the jet interpretation, one could search for other possible
solutions by considering the temporal and spectral relations in various
afterglow models 
(e.g. M\'esz\'aros, Rees \& Wijers 1998; Sari, Piran \& Narayan 1998; 
Chevalier \& Li 2000; Zhang \& M\'esz\'aros 2004). 
The most straightforward model is within the framework of the standard
isotropic wind model (Chevalier \& Li 2000).
The first cluster of the data corresponds to the $\nu_c<\nu_X<\nu_m$
regime, in which the temporal index $-1/4$ and the photon spectral
index $-3/2$ are expected. The second and the third clusters of the
data correspond to the regime of $\nu_m < \nu_X < \nu_c$, in which the
temporal index $-(3p-1)/4$ and the photon spectral index $-1-(p-1)/2$ are
expected. All these are consistent with the data for $p\sim 2.1$. In
this interpretation, one needs to assume that both $\nu_m$ and $\nu_c$
cross the X--ray band during the gap between the first two clusters of
data and that the frequencies switch the order. This could be achieved
with a small wind parameter (e.g. $A_*$ in the range of $0.01-0.001$).
One caveat is that in the wind model $\nu_c \propto t^{1/2}$, so that
the time interval of the gap is not long enough for $\nu_c$ to
completely cross the band. Nonetheless, the spectral slope in the
second cluster is slightly steeper than the other two, which might be
still consistent with the data if one introduces an evolving cooling
break near the high energy edge of the band during the epoch of the
second snapshot. Introducing a slightly steeper density profile
(larger than $r^{-2}$ for the wind case) could further alleviate the problem.
Furthermore, if the electron equipartition factor $\epsilon_B$ drops
during the temporal gap, this would speed up the $\nu_c$ crossing time scale, 
making the model more consistent with the data.

Finally, we note also that a similar behaviour has been observed in optical
light curve GRB021004 even if with a slightly longer break time ($\sim 0.1$ d,
Fox et al. 2003). This transition has been interpreted by Li \& Chevalier
(2003) as a fast to slow cooling transition.  

\section{Conclusion}

Swift is exploring for the first time the early stages of GRB afterglow
decays. We detect a clear early temporal break in the X--ray afterglow
of GRB050128, with the spectral indices not changing around the break.
The data could be argued to be consistent with either a jet model or a
wind model. 
The jet model requires a flat electron spectrum and an assumed
spectral domain change within the temporal gap between the first two
snap shot observations. If this is true, we may have detected the
earliest jet break so far.
The wind model requires a (relatively) low ambient density and possibly an
evolution of the $\epsilon_B$ parameter. 
We note that in this last case, passages from fast to slow cooling 
regimes might have remained hidden in the great majority of GRB afterglows 
if they are characterized by such a tenuous environment, due to the late 
times at which they were observed.

The early detection of the X--ray afterglow, coupled with the initial
flat decay, allows us to estimate its fluence $\mathcal{F} = \int
F(t){\rm d}t$. In fact, given the value of $\alpha_1$ and $\alpha_2$,
$\mathcal{F}$ is not very sensitive to the start time $t_0$ of the
afterglow and is dominated by the flux at the break time
$t_b$. The fluence between 108 s and 73000~s is $(7\pm2) \times 10^{-7}$ erg
cm$^{-2}$ (15--350 keV), while extrapolating from $t = 0$ 
to $+\infty$ it is $\mathcal{F} = 9^{+5}_{-3} \times 10^{-7}$ erg
cm$^{-2}$. These values amount to $15\%$ and $20\%$ of the prompt fluence
in the same energy band. Since prompt and afterglow spectra are similar we
might expect a relatively small difference in the bolometric correction. This
is the first determination of the ratio between GRB proper and early afterglow
energetics.  

\begin{acknowledgements}
This work is supported at OAB by funding from ASI on grant number I/R/039/04,
at Penn State by NASA contract NAS5-00136 and at the University of Leicester
by PPARC on grant numbers PPA/G/S/00524 and PPA/Z/S/2003/00507.  We gratefully
acknowledge the contributions of dozens of members of the XRT and UVOT team at
OAB, PSU, UL, GSFC, ASDC, MSSL and our subcontractors, who helped make this
instrument possible. 
\end{acknowledgements}

\newpage

\begin{figure*}[htb]
\psfig{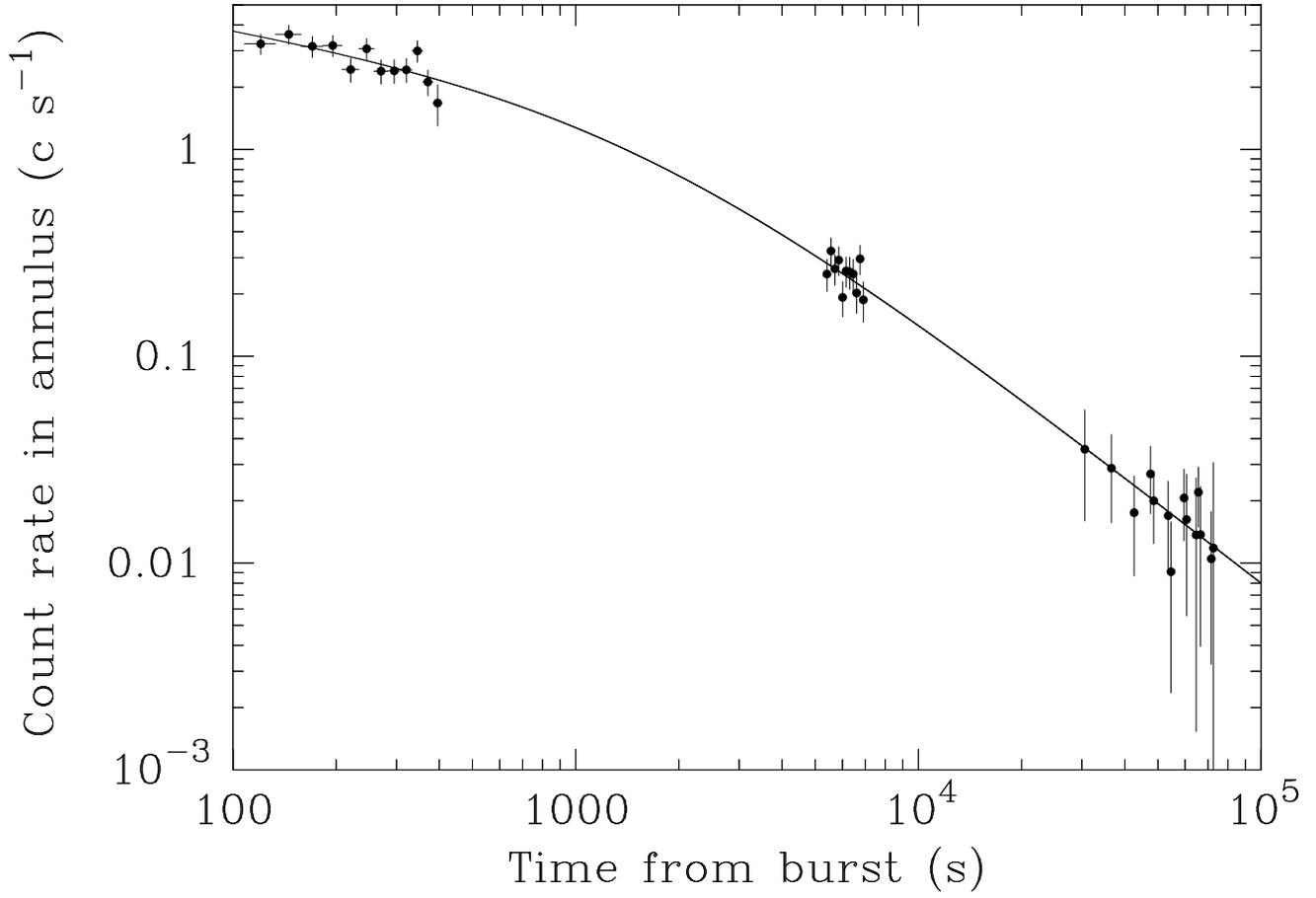}
\caption{XRT 0.2--10 keV light curve extracted from an annular region centered
on GRB050128. The continuous line represents the fit with two power laws
smoothly joined. The conversion factor to translate the count rate into a
0.2--10 keV unabsorbed flux is $8.7\times 10^{-11}$ erg cm$^{-2}$
counts$^{-1}$ (for a freely varying absorbed power law model, see text).}
\label{fig:lc}
\end{figure*}

\begin{figure*}
\begin{center}
\psfig{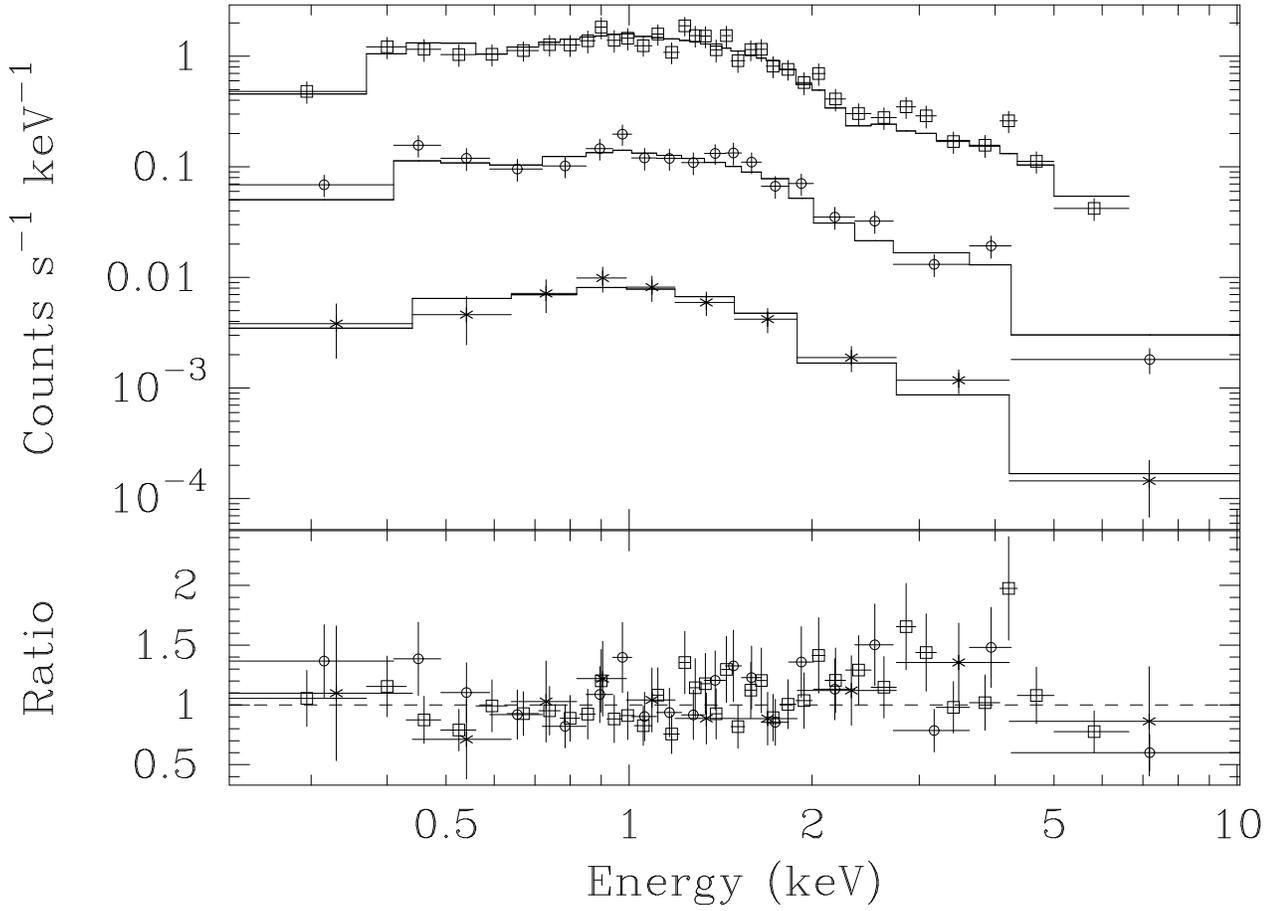}
\end{center}
\caption{XRT 0.2--10 keV energy spectrum of GRB050128. In the upper panel are
plotted the spectra of the three snapshot observations described in the text
(1: squares, 2: circles, other: stars) fit with an (freely) absorbed power law model. 
In the lower panel there are the residuals from the same power law fit to all the data.}
\label{fig:spe}
\end{figure*}

\begin{figure*}
\begin{center}
\psfig{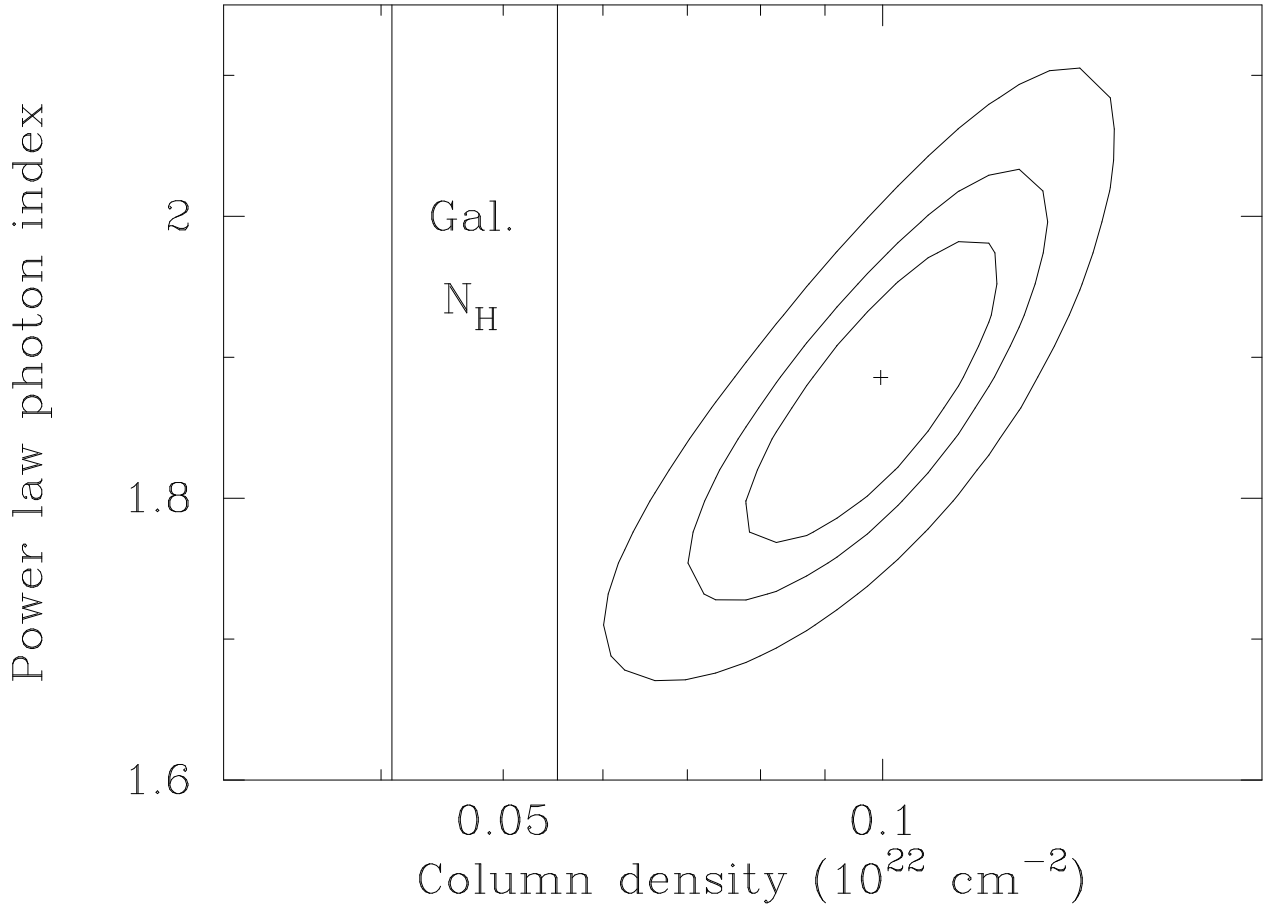}
\end{center}
\caption{Contour plot of the column density versus the power law photon index
for the X--ray spectrum of GRB050128. Contours refer to 1, 2 and $3\sigma$
confidence level. At the left of the contour plot the Galactic column density
interval is drawn centered on the value of $4.8\times 10^{20}\cmdue$ and with
a $15\%$ uncertainty.}
\label{fig:cont}
\end{figure*}

\end{document}